\documentclass[twocolumn,aps,tightenlines,floatfix,showpacs]{revtex4}

\usepackage{graphics}
\usepackage{bm}
\usepackage{amsmath}
\usepackage{amssymb}

\begin{document}

\title{Predicting narrow states in the spectrum of a nucleus beyond the 
proton drip line}
                                                                                
\author{L. Canton$^{(1)}$}
\email{luciano.canton@pd.infn.it}
\author{G. Pisent$^{(1)}$}
\email{gualtiero.pisent@pd.infn.it}
\author{J. P. Svenne$^{(2)}$}
\email{svenne@physics.umanitoba.ca}
\author{K. Amos$^{(3)}$}
\email{amos@physics.unimelb.edu.au}
\author{S. Karataglidis$^{(3)}$}
\email{kara@physics.unimelb.edu.au}
                                                                                
\affiliation{$^{(1)}$ Istituto  Nazionale  di  Fisica  Nucleare,
   Sezione  di Padova, e\\  Dipartimento di Fisica  dell'Universit\`a
   di Padova, via Marzolo 8, Padova I-35131, Italia,}
\affiliation{$^{(2)}$ Department  of  Physics  and Astronomy,
   University  of Manitoba, and Winnipeg  Institute for   Theoretical
   Physics, Winnipeg, Manitoba, Canada R3T 2N2}
\affiliation{$^{(3)}$ School of Physics, The University of Melbourne,
   Victoria 3010, Australia}
                                                                                
\date{\today}

\begin{abstract}
Properties of particle-unstable nuclei lying beyond the proton drip 
line can be ascertained by considering those (usually known) properties
of its mirror neutron-rich system. We have used a multi-channel algebraic
scattering theory to map the known properties of the neutron-${}^{14}$C
system to those of the proton-${}^{14}$O one from which we deduce that 
the particle-unstable ${}^{15}$F will have a spectrum of two low lying 
broad resonances of positive parity and, at higher excitation, three 
narrow negative parity ones. A key feature is to use coupling to 
Pauli-hindered states in the target. 
\end{abstract}
                                                                                
\maketitle


There is much current interest in the properties of, and reactions with, 
nuclei that lie out of the valley of stability. The masses of thousands of 
such nuclei that lie between the nucleon drip lines are known as are some
spectral properties of those that can be formed with sufficient intensity
for a radioactive ion beam (RIB) to be made. Besides the inherent interest
in studying the properties of weakly-bound many-nucleon systems, 
these radioactive nuclei are crucial in current investigations of 
energy and mass production in astrophysics.

Little is known about nuclear systems at and beyond the drip lines. Those  
particle-unstable systems are difficult to study as they may only be formed
during nuclear reaction processes. However, as they can have much
influence on  nucleogenesis in astrophysical calculations, their properties
need be understood. Recently, data have been obtained~\cite{Go04,Gu05} from 
elastic scattering of radioactive ${}^{14}$O ions from hydrogen which reveal
two states in the proton-rich nucleus ${}^{15}$F; a nucleus that lies  outside 
of the proton drip line.  Those data indicated that besides the resonant ground 
state, ${}^{15}$F had a narrower first excited resonant state 1.3 MeV above 
the (broad resonance) ground.  Herein we report on our analysis of  those data 
and predict that there should be even narrower resonances in ${}^{15}$F
lying in an energy range just above the limit of the reported data.  

As method of analysis we use the multi-channel algebraic scattering 
theory (MCAS)~\cite{Am03}. It has the distinctive capacity
to embrace in the scattering equations single-particle aspects, 
collective-type coupled-channel dynamics, and the Pauli principle. 
The Pauli principle is taken into account using the 
Orthogonalizing-Pseudo-Potential (OPP) method. 
Past studies~\cite{Am03,Ca05,Pi05} used that OPP scheme to deal only with 
Pauli-blocked and Pauli-allowed states.  In this letter we use the OPP scheme
to consider also Pauli-hindered states, namely states where the 
Pauli-blocking is partially relaxed (Pauli-hindrance).
With that new feature, and with the instructive property of 
considering results in the limit of zero deformation~\cite{Pi05}, 
our analyses of the $p$+${}^{15}$O  system and of
its mirror, the $n$+${}^{14}$C system, infers new spectroscopy of the 
very exotic nucleus, ${}^{15}$F.

The concept  of Pauli-hindrance relates to  levels that are 
neither Pauli-forbidden nor Pauli-allowed but are somewhere in between. 
This concept naturally arises in cluster-dynamics formulations 
based, for example, on the Resonating Group Method (RGM). 
Therein, such conditions can be studied 
in detail, even analytically, starting from the properties of the 
eigenvalues of the RGM norm kernel~\cite{Sc78}. The technique based
on the introduction of the Orthogonalizing Pseudo-Potential (OPP) 
method, which we have adopted and generalized to multichannel dynamics 
in the MCAS formulation, is particularly suited for treating such 
intermediate situations.  For reference,
Pauli-allowed states relate to zero coupling in the OPP term and 
complete Pauli-blocking is the limit of infinite OPP couplings. 
In practice, blocking-effects can be obtained numerically by having 
large (of order GeV) values to the OPP couplings, while for 
Pauli hindrance couplings of the order of a few MeV are required in 
the strength of the corresponding OPP term. 
In our current formulation of the MCAS approach, we had to include this 
concept of Pauli-hindrance in the OPP scheme to deal with breaking 
effects in shell closures, particularly of $0p_{\frac{1}{2}}$ proton orbits, 
which is a physical phenomenon to be expected in weakly-bound 
light exotic nuclei.

Shell-closure aspects represent not only a fundamental question 
in current research in nuclear structure and reactions involving exotic 
nuclei, but are also of great relevance for atomic and 
molecular physics in general. In addition, breaking signals in the 
full occupancy of deep and  well-packed orbits are the subject of a new 
proposal of studies in atomic physics~\cite{Mi05}, specifically 
regarding possible upper limits in the violation of the Pauli 
principle (VIP). We stress, in this respect, that the 
shell-breaking phenomena in weakly-bound (or unbound) nuclei 
that we consider in this Letter, and the related concept of Pauli-hindrance,  
are entirely consistent with the validity of the Pauli principle.

Use of the MCAS approach in the analysis of scattering data (of nucleons
and nuclei) has the advantage that such nontrivial effects of Pauli principle
can be incorporated with the OPP method in the multichannel scattering
equations. Sturmian expansions of the nuclear interactions
are used to obtain an algebraic form for the multichannel S-matrices.
The method treats bound as well as continuum regimes of the compound 
system equally, and incorporates a resonance-finding procedure by which 
all bound states and all resonances up to the limit energy considered will 
be defined (spin, parity, centroid energy, and width). 
That is so no matter how narrow or broad any resonance may be. 
Importantly, use of the OPP method in the construction of the Sturmian 
functions ensures that the Pauli principle is not violated even when 
a collective-model prescription of the nucleon-nucleus interactions 
is used. 

Low-excitation bound states and resonances in the spectra of nuclei in 
the mass region A $\sim$ 13-31 include many that are expected to be due
to weak coupling of a nucleon in the $s-d$ shell to the (A-1) nucleon
core. Such has  been seen in the spectrum of ${}^{15}$C with the ground
and first excited  states being bound and having spin-parities of
$\frac{1}{2}^+$ and  $\frac{5}{2}^+$ and with energies lying below the
$n$+${}^{14}$C threshold by 1.218 and 0.478 MeV
respectively~\cite{Aj91}.  On the other hand, the observed two
resonances in ${}^{15}$F  are centered about 1.47 and 2.78  MeV above
the $p$+${}^{14}$O threshold. However they match the spin-parity  values
of the two bound states in ${}^{15}$C and are considered their
analogues. Hence we consider the $n$+${}^{14}$C system and the states in
${}^{15}$C first and match the result by adding a Coulomb field in the
calculations to specify the spectrum and scattering cross section for
$p$+${}^{14}$O. That spectroscopy is determined from MCAS evaluations,
input to which are interaction potentials for the channels coupled in
the systems. Both mass 14 nuclei have a $0^+$ ground state and then a
cluster of excited states some 6 MeV away. In that cluster there are  a
second $0_2^+$, a $2^+$, a $1^-$ and a $3^-$ state. Of those, for
simplicity in calculations, we consider coupling to the ground only with
the $0_2^+$ and the $2^+$ states. (We have also considered 
alternative couplings to other excited states, but results then are 
definitively inferior.)
The quadrupole coupling strength is taken as $\beta_2 = -0.5$  which
is similar to the value used for the $n$+${}^{12}$C  system in a previous 
MCAS analysis~\cite{Am03,Ca05,Pi05}. 

We presume that the ground states are described dominantly by two holes
in an otherwise closed ${}^{16}$O. Thus for ${}^{14}$C (neutrons) and 
${}^{14}$O (protons), the $0s_{\frac{1}{2}}, 0p_{\frac{3}{2}}$, and 
$0p_{\frac{1}{2}}$ relevant nucleon orbits in the ground states are 
considered full. For the ground-state channels in the nucleon-nucleus
systems, then, those orbits are Pauli blocked while all other orbits are
treated as Pauli allowed. However, we presume that the excited states
are dominated by 2 particle $-$ 4 hole (and higher) configurations with
the occupancies of the $0p_{\frac{1}{2}}$ orbits most affected. Thus
we treat that orbit, for the relevant nucleon type and in the channels 
involving the excited states, as Pauli hindered. All such Pauli principle
effects are generated using the OPP scheme by which the Sturmians
are orthogonal to any Pauli-blocked state and affected by any that are
Pauli hindered.  Those Sturmians
are used to expand the interaction matrix of potentials and then 
the scattering matrices.

The Sturmians are solutions of homogeneous Schr\"odinger equations for the  
matrix of potentials. In coordinate space, if those potentials are designated 
by local forms $V_{cc'}(r) \delta(r-r')$, the OPP method uses Sturmians 
that are solutions for nonlocal potentials
\begin{equation}
\mathcal{V}_{cc'}(r,r')  =  V_{cc'}(r)\delta(r-r')  +  \lambda_c
    A_c(r) A_c(r')\delta_{cc'} ,
\label{OPPeq}
\end{equation}
where $A_c(r)$ is the radial part of the single-particle bound-state wave 
function in channel $c$ spanning the phase space excluded by the Pauli 
principle.  The channel indices $c$ designate all relevant quantum numbers.
The OPP method for treating Pauli-blocked state effects, holds 
in the limit $\lambda_c \to \infty$, but use of $\lambda_c = 1000$~MeV 
suffices.  For Pauli-allowed states, of course, $\lambda_c = 0$.  But for 
Pauli-hindered states specific values of 1000~MeV $>> \lambda_c > 0$ 
are required. 
Those strengths ($\lambda_c$) are presently treated as parameters though  
microscopic or cluster models of structure could be used to generate them.

We use the same collective model prescription for the matrix of interaction
potentials that we have used previously~\cite{Am03,Ca05,Pi05} but in this
case with the mix of central (0), spin-orbit (so), and $l^2$ (ll)
deformed potential terms,
\begin{equation}
V_{c'c}(r) =  V_0 v^{(0)}_{cc'}(r,\beta_2)
+ V_{so} v^{(so)}_{cc'}(r,\beta_2) +  V_{ll} v^{(ll)}_{cc'}(r,\beta_2)\ ,
\label{Vmat}
\end{equation}
where quadrupole deformation is taken to second order~\cite{Am03} and
the basic functional form is that of a Woods-Saxon  
with an undeformed radius of 3.1 fm and a diffuseness of 0.65 fm.
The successful calculations of the neutron-${}^{14}$C system required
potential strengths of $V_0 = - 45.0$ MeV, of $V_{so} = 7.0$ MeV, and of 
$V_{ll} =  0.42$ MeV.  The very same interaction was used to determine the 
positive and negative parity results so that the only parity dependence 
arises from use of the OPP term in Eq.~(\ref{OPPeq}). 
The Coulomb radius used in the $p$+${}^{14}$O calculations was 3.1 fm and, 
with respect to the $n$+${}^{14}$C system, it was necessary to reduce the 
central potential strength $V_0$ slightly, to -44.2 MeV.

\begin{figure}[t]
\scalebox{0.5}{\includegraphics{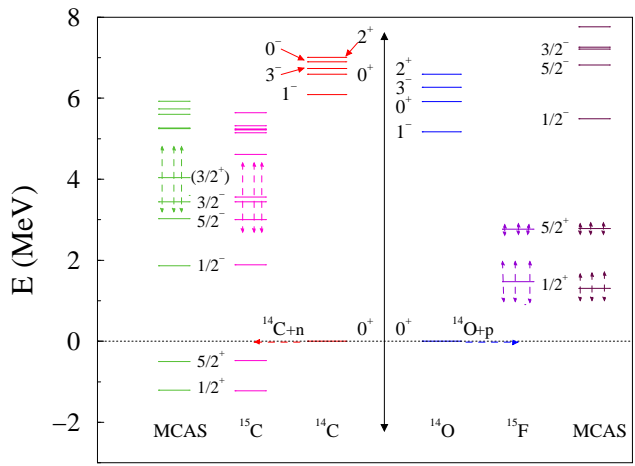}}
\caption{\label{Fig1}(Color online)
Low energy spectra of ${}^{14,15}$C and of ${}^{14,15}$O and of the results
from our MCAS calculations. The zero of the energy scale is set to that
of the relevant mass 14 ground states.}
\end{figure}

The spectra, known and calculated using MCAS, are shown in Fig.~\ref{Fig1}.
The specific cases are as indicated in the diagram.
Considering the ${}^{14,15}$C results first, the excited states of the 
target ${}^{14}$C are clustered and well separated by some 6~MeV from
the ground. The spectrum of ${}^{15}$C has two bound states of
spin-parities $\frac{1}{2}^+$ (ground) and $\frac{5}{2}^+$ and which are
dominantly described by a single $s$-$d$ shell neutron on the ${}^{14}$C 
ground state. Then there are three quite narrow resonances; all having
negative parity which lie within the spread of a broad $\frac{3}{2}^+$
resonant state. That broad $\frac{3}{2}^+$ was seen very clearly in
the cross section from a measurement~\cite{Da85} of the ${}^{14}C(d,p)$ 
reaction. The MCAS result matches all of those features well.
In the zero deformation limit ($\beta_2 \to 0$), the MCAS results reveal 
that the bound ($\frac{1}{2}^+$ and $\frac{5}{2}^+$) and resonant 
$\frac{3}{2}^+$ states are due to the coupling of an $1s_{\frac{1}{2}}$, of a
$0d_{\frac{5}{2}}$ and of a $0d_{\frac{3}{2}}$ neutron to the ground
state of ${}^{14}$C. It is noteworthy that there are no other bound states
and especially of negative parity. Such would occur if in the $n$+${}^{14}$C
system the $0p_{\frac{1}{2}}$ neutron orbit were not Pauli blocked. The
negative parity states have as their progenitor a $0p_{\frac{1}{2}}$
coupled to the $0_2^+$ state (for the $\frac{1}{2}^-$ state) and to the
$2^+$ state (for the $\frac{3}{2}^-$ and $\frac{5}{2}^-$ states). To find
these states at this excitation in ${}^{15}$C required that the 
Pauli-hindrance of the neutron $0p_{\frac{1}{2}}$ orbit in 
the $0^+_2$ and $2^+$ states of ${}^{14}$C target be generated 
with $\lambda_c(0p_{\frac{1}{2}})$ values of 3.11 and 3.87 MeV, respectively.

Scattering cross-section results are shown in Fig.2. In the top panel
the cross sections from ${}^{14}$O scattering from hydrogen (in inverse 
scattering of protons from ${}^{14}$O) at 180$^\circ$ in the center of mass
are given. Therein our MCAS result (solid curve) is 
compared with the recent data of both Goldberg {\it et al.}~\cite{Go04} 
(open circles) and Guo {\it et al.}~\cite{Gu05} (filled squares). 
The Guo data were in arbitrary
units and so we normalized them to the $\frac{5}{2}^+$ resonance values of
Ref.~\cite{Go04}. Though the more recent experiment obtained results to 6 MeV, 
the authors indicate that such are reliable to about 5 MeV. 
\begin{figure}[t]
\scalebox{0.5}{\includegraphics{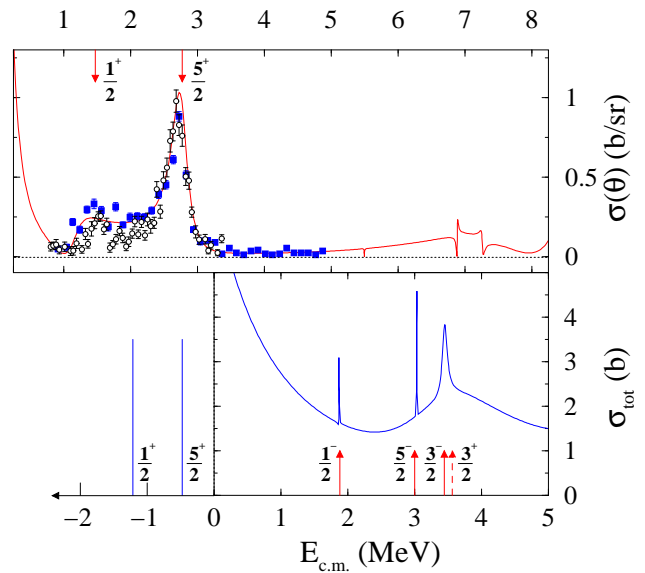}}
\caption{\label{Fig2}(Color online)
The elastic cross sections from scattering of ${}^{14}$O ions from hydrogen
at 180$^\circ$ in the center of mass (top) and our predicted total cross 
section for the scattering of neutrons from ${}^{14}$C (bottom). In both cases
the known spectral values are indicated by the arrows.}
\end{figure}
In the bottom panel of Fig.~\ref{Fig2} we show our prediction of the 
total scattering cross section of neutrons from ${}^{14}$C for energies
to 5 MeV. The zero of the energy scale has been placed to optimally match 
the $\frac{5}{2}^+$ bound state in ${}^{14}$C to the centroid of
the analogous resonance state in ${}^{15}$F.

The experimental values~\cite{Aj91} of states in the two mass 15 systems are
indicated by the arrows with the relevant spin-parities given alongside.

\begin{figure}[t]
\scalebox{0.5}{\includegraphics{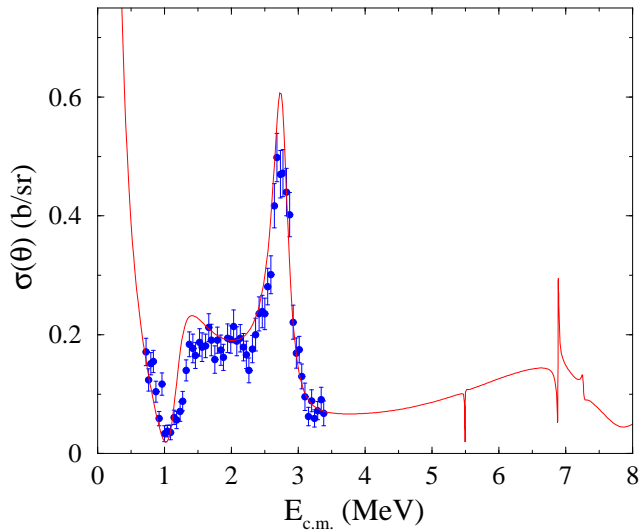}}
\caption{\label{Fig3}(Color online)
The elastic cross sections from scattering of ${}^{14}$O ions from hydrogen
at 147$^\circ$ in the center of mass. The data was taken from 
Ref.~\cite{Go04}.}
\end{figure}

Consider the results for the neutron total cross section from ${}^{14}$C.
That cross section has four obvious resonances, three quite narrow
(the negative parity resonances) and one, a $\frac{3}{2}^+$ resonance, 
very broad. That broad resonance agrees with one such found in the
cross section from the stripping reaction, ${}^{14}$C$(d,p)$~\cite{Da85}.
All of these features have a partner in the 180$^\circ$ cross section for
the $p$+${}^{14}$O system that is shown in the top panel.
Clearly the MCAS fit to the available data is good and as good as has been
found with other analyses~\cite{Go04,Ba05}.
Noteworthy is that the ground state of the particle-unstable
${}^{15}$F is an $s$-wave resonance. That is so only because of the Coulomb
barrier in the $p$+${}^{14}$O system. Without the Coulomb barrier there
would be no $s$-wave resonance, only a virtual bound state~\cite{Ta72}. 
That criticality was the reason we needed a small reduction in the central
interaction strength (of but 0.8 MeV) to locate this resonance properly.
Otherwise the interactions used were exactly those determined by our
study of the $n$+${}^{14}$C system. The two bound states found for 
${}^{15}$C have  become resonances and are of single-particle-like 
nature.  There are other resonance features in our
calculated results lying just above the highest energy at which experimental
results are known to date. These have negative parities and are analogues of
the negative parity resonances seen in ${}^{15}$C. Thus the origin of 
these new, narrow  negative parity resonances in ${}^{15}$F differ 
from those of the observed  low-lying ones. They are compound resonances 
and, as with those identified in ${}^{15}$C, are due to the Pauli-hindrance
of the proton $0p_{\frac{1}{2}}$ orbit in the $0_2^+$ and $2^+$ 
excited states of ${}^{14}$O. Finally, we note that these new 
resonances persist and are relatively more noticeable in cross-sections 
at other scattering angles. As an example, we show in Fig.~\ref{Fig3}, 
results from our MCAS calculation compared with data~\cite{Go04} 
taken at 147$^\circ$.
Again the two low lying, broad resonances are predicted well (location, 
width and magnitude) and now the higher, narrow, negative parity resonances
are clearly seen to reside on a broad ($\frac{3}{2}^+$) resonance. That 
broad resonance is the analogue of that seen in the ${}^{14}$C$(d,p)$
experiment~\cite{Da85}.

In conclusion, the MCAS approach has been used with mirror mass 15
systems to define the spectroscopy  of the particle-unstable nucleus,
${}^{15}$F. The procedure involved  first making an analysis of the
neutron-mirror mass ${}^{15}$C system for which experimental information
is known. Crucial to the description of the experimental spectrum was
the concept of Pauli-hindrance of single-particle orbits  coupled to the  
collective $0_2^+$ and $2^+$ excitations in the mass  14 nuclei. It
leads to the correct description of the observed  three low-lying
negative-parity resonances. Then, by incorporating  Coulomb  
interactions, the same nuclear force was used to analyze the
proton-${}^{14}$O  case and thus to predict the spectroscopy of
${}^{15}$F up to 8 MeV excitation. We clearly see three
narrow negative-parity resonances in the calculated
cross section. This demands further experiments to test the theoretical
interpretation.

The scheme we have used may be repeated to estimate 
spectroscopy of other nuclei that are just outside of the
proton   drip line given that the numbers of neutron-rich isotopes
within the neutron   drip line usually exceed those on the proton-rich
side. Thus the mirror system  against which the proton-rich, unstable,
system spectroscopy is to be compared  will not be particle unstable and
may possibly have experimentally known  and detailed  properties.

\begin{acknowledgments}
  This research was supported by a grant from the Australian Research
Council, by a merit award with the Australian Partners for   Advanced
Computing, by the Italian MIUR-PRIN Project      ``Fisica Teorica del
Nucleo e dei Sistemi a Pi\`u Corpi'', and by the Natural Sciences and
Engineering Research Council (NSERC), Canada. KA and JPS also thank
the INFN, sezione di Padova, and the Universit\`a di Padova for financial 
support of their visits to Padova for collaboration. 
\end{acknowledgments}

\bibliography{PRL-mass15}
\end{document}